# Structure analysis of Zn-Mg-Ho icosahedral quasicrystal by modified Rietveld method using ellipsoid and sphere windows

Tsutomu Ishimasa[a,*], Kei Oyamada[a], Yasuomi Arichika[b], Eiji Nishibori[c], Masaki Takata[c,d], Makoto Sakata[c] and Kenichi Kato[d]

[a] *Division of Applied Physics, Graduate School of Engineering, Hokkaido Univ., Kita-ku, Sapporo 060-8628, Japan*
[b] *Division of Nuclear Engineering, Graduate School of Engineering, Nagoya Univ., Chikusa-ku, Nagoya 464-8603, Japan*
[c] *Division of Applied Physics, Graduate School of Engineering, Nagoya Univ., Chikusa-ku, Nagoya 464-8603, Japan*
[d] *Japan Synchrotron Radiation Research Institute (JASRI), 1-1-1 Kouto, Mikazuki-cho, Sayo-gun, Hyogo 679-5198, Japan*

Abstract

A structure analysis of Zn-Mg-Ho icosahedral quasicrystal was carried out by the powder X-ray diffraction method using synchrotron radiation ($\lambda=0.73490$ Å) at SPring-8. The intensity distribution was analyzed by the Rietveld method modified for an icosahedral quasicrystal, in which simplified models were assumed: ellipsoid and spherical windows were assumed at five crystallographic sites in a F-type hypercubic unit cell. The analysis revealed the presence of an almost perfect Penrose tiling with edge length 5.20 Å. The vertices are occupied alternatively by Zn and Mg, and almost all of the edge centers of the Penrose rhombohedra are occupied by 0.8Zn and 0.2Mg. Ho and Mg atoms tend to be present on the body diagonal of the prolate rhombohedra. Good agreement between the measured and calculated intensity distribution using the simplified model suggests the applicability and the limitation of structure analysis using the powder X-ray diffraction method.

* Corresponding author:   Tsutomu Ishimasa
Division of Applied Physics, Graduate School of Engineering, Hokkaido Univ., Kita-ku, Sapporo 060-8628, Japan
Telephone:   +81-11-706-6643
Fax:   +81-11-716-6175
e-mail address:   ishimasa@eng.hokudai.ac.jp

1. Introduction

The powder X-ray diffraction technique using synchrotron radiation may have the potential ability to analyze an atomic arrangement in a quasicrystal. If one remembers that very many reflections are distributed in the reciprocal space of a quasicrystal and we do not know exactly the positions where we should measure, one can imagine that the powder X-ray diffraction technique can be complementary to the single-quasicrystal diffraction technique, even though the directional information is lost in this technique. In this paper, we will report the applicability and the limitation of the modified Rietveld method to the structure analysis of the Zn-Mg-Ho icosahedral quasicrystal, assuming a simplified model with ellipsoidal and spherical windows. The Zn-Mg-Ho quasicrystal is a member of a series of icosahedral quasicrystals consisting of Zn, Mg and rare-earth metals [1,2]. These quasicrystals have the following interesting features: they form as equilibrium phases in the cases of Y, Dy, Ho and Er, and exhibit short-range magnetic ordering reflecting the icosahedral symmetry [3]. From the structural point of view, they may be classified into Bergman-type with F-type atomic ordering [4,5]. The Ho atom has a larger number of electrons than Zn and Mg, indicating that good contrast is expected for the Ho in the X-ray diffraction analysis. Actually, the preceding studies have indicated that the F-type ordering is related to the arrangement of the Ho atoms [4,5,6]. The purpose of the present work is to find a simple structure model of the Zn-Mg-Ho icosahedral quasicrystal that preserves essential structural properties of this quasicrystal.

2. Experimental procedure

An alloy ingot with a nominal composition of $Zn_{59.0}Mg_{31.0}Ho_{10.0}$ was synthesized by the method described in an earlier paper [7]. The heat treatments applied in this work consist of two parts, melting and annealing. The weighed materials were packaged in molybdenum foil to prevent both evaporation and reaction with the silica ampoule, and were then sealed in an argon atmosphere of 200 Torr in the silica ampoule after evacuation to $4 \times 10^{-6}$ Torr. They were melted at 774 °C for 3 h and then quenched in water without breaking the ampoule. The alloy ingot thus made was annealed at 560 °C for 50 h for homogenization, and successively annealed at 504 °C for 54 h in order to increase the atomic ordering [5]. After the annealing process the ampoule was quenched in water. The weight loss during the heat treatments was measured to be less than 0.01 %, and thus the deviation of the alloy composition from the nominal one was regarded as very small. The mass density of the Zn-Mg-Ho quasicrystal was measured to be 5.64±0.05 g/cm$^3$ by Archimedes' method.

Powder X-ray diffraction experiments was carried out at beam line BL02B2 at SPring-8. A Debye-Scherrer camera with a diameter of 286.5 mm, which was equipped with an imaging plate, was used for the diffraction experiment. The powder specimen was put in a fine glass capillary tube, and the tube was rotated at the center of the camera. The resolution in 2θ was 0.01 degree. The diffraction intensities measured at 4301 points were used for the Rietveld analysis, which correspond to a 2θ range between 7 and 50 degrees. The wavelength of the X-ray was estimated to be 0.73490 Å from the lattice parameter of standard $CeO_2$, a=5.4113 Å [8].

3. Rietveld method modified for an icosahedral quasicrystal

The so-called Rietveld method [9] is a powder X-ray diffraction analysis of the crystal structure using intensity profile fitting. The intensity $Y$ at 2θ is expressed by the equation (1),

$$Y(2\theta) = S \sum_k LP(\theta_k) M_k |F_k|^2 P(2\theta - 2\theta_k) + BG(2\theta)  \quad\quad \text{--- (1)}$$

where $S$, $LP$, $\theta_k$, $M_k$, $F_k$, $P$ and $BG$ denote, respectively, the scale factor, the Lorentz polarization factor, the Bragg angle of the $k$-th reflection, its multiplicity, the structure factor, the peak shape and the background intensity. The intensity $Y(2\theta)$ was obtained by summing up the contributions of nearby reflections. The ordinary computer program of the Rietveld method assumes periodicity in the 3-dimensional space, and must be modified for the analysis of a quasicrystal.

The structure factor of an icosahedral quasicrystal is calculated on the assumption that the structure of the quasicrystal can be regarded as a 3-dimensional section of a 6-dimensional hypercubic structure [10,11,12]. The structure factor $F(\boldsymbol{g}_{//})$ is calculated from equation (2) assuming periodicity in the 6-dimensional space,

$$F(\boldsymbol{g}_{//}) = \sum_n f_n(|\boldsymbol{g}_{//}|) \left\{ \int W_n(\boldsymbol{r}_\perp) \exp(2\pi i \boldsymbol{g}_\perp \cdot \boldsymbol{r}_\perp) d\boldsymbol{r}_\perp \right\} \exp(2\pi i \boldsymbol{g} \cdot \boldsymbol{C}_n) \exp(-(B_{//}|\boldsymbol{g}_{//}|^2 + B_\perp |\boldsymbol{g}_\perp|^2)/4)  \quad\quad \text{--- (2)}$$

where $\boldsymbol{g}_{//}$, $\boldsymbol{g}_\perp$, $f_n$, $W_n$, $\boldsymbol{C}_n$, $B_{//}$ and $B_\perp$ denote, respectively, the reflection vectors in the physical and perpendicular spaces, the atomic scattering amplitude, the window function expressing the shape of the

*n*-th atomic motif (or window), the position vector of the window, the isotropic *B* factors in Debye-Waller factors in the physical and the perpendicular spaces. The values of the atomic scattering amplitudes described in Ref. 13 are used. The anomalous dispersion terms were included, which were calculated by the software CROMER based on the Cromer-Liberman method [14]. A unique set of *B* factors were used for all of the windows in the present analysis. Since the type of hypercubic lattice in the 6-dimensional space is known in the present case, the goal of this study is to determine the positions and the shapes of windows as well as the corresponding atom type for each window.

The simplified model presented in Table 1 is assumed referring to the results of former studies [4,5,6]. The shapes of the windows, spherical and ellipsoidal, satisfy the required restrictions on the site symmetry. Corresponding to the F-symmetry of the Zn-Mg-Ho icosahedral quasicrystal, the sites at Node1 and Node2 are differentiated, and similarly BC1 and BC2. The site symmetry, $C_{5v}$, of the edge center 1/4[1 0 0 0 0 0] is lower than the icosahedral symmetry, $Y_h$. We assumed the ellipsoid at the edge-center position, EC, which can be shifted along the 5-fold axis in the perpendicular space as indicated by Takakura et al. [6]. Note that the sites e.E and o.E in Ref. 6 are equivalent to each other, and correspond to EC in Table 1. All fitting parameters used in the Rietveld analysis are listed in Table 2. The first 16 parameters are those describing the quasicrystal structure, and the remaining 20 are related to the scale factor, the correction of peak position and the peak shape including asymmetry and background.

To search the 6-dimensional model, we have used the following assumptions in addition to the shapes of the windows:
(1) Perfect F-type ordering.
(2) Windows related to unique atom types, except for EC. In particular, almost all of the Ho atoms belong to the window located at BC2 [4,5,6].
(3) Adjustment of alloy composition by controlling the composition of EC, which includes Zn and Mg.
The third assumption is because the number of EC windows is 12 times larger than that of the other windows, and the alloy composition depends strongly on the composition of the EC. The computer program itself is applicable to the multiple shell windows of spheres for Node1, Node2, BC1 and BC2. However, in reality, a good fit was obtained by using simple spheres at these four sites. The structure factors of 3005 independent reflections were used to calculate the intensity distribution, which satisfy the following conditions on the reflection vectors $\boldsymbol{g}_{//}$ and $\boldsymbol{g}_{\perp}$; $0.037 < g_{//} < 1.256$ Å$^{-1}$ and $g_{\perp} < 0.3$ Å$^{-1}$. Several assignments of Zn and Mg to Node1, Node2 and BC1 were tested as the initial structure models. In these models, 0.7Zn+0.3Mg was assumed for EC. In the subsequent steps, the composition of EC was changed manually. The search was repeated until models were obtained that exhibit good agreement with the experimental intensity distribution and satisfy the conditions on the weighted reliable factor $R_{wp}$, the alloy composition, the mass density and the occurrence of short interatomic distances.

4. Results and Discussion

The powder X-ray diffraction pattern of the $Zn_{59.0}Mg_{31.0}Ho_{10.0}$ alloy is presented in Fig. 1(a). More than 185 diffraction peaks were identified as those of the icosahedral quasicrystal, which were indexed by six integers either all even or all odd. This reflection condition indicates the presence of F-type ordering. The peak is very sharp and the peak widths, $\Delta g_{//} = \Delta q/2\pi$, range between 1.2 and 1.8x10$^{-3}$ Å$^{-1}$, measured as the full width at half maximum. There was no clear dependence of the peak widths on the magnitude of $g_{\perp}$, but a weak dependence on $g_{//}$ was observed, which may be caused by the preparation of the powder specimen. Careful examination revealed that 16 very weak peaks were not assigned to the icosahedral quasicrystal. The strongest unidentified peak exists at $2\theta =$ 17.12° near the base of the $22\bar{2}2\bar{2}4$ reflection. This peak has an integrated intensity of 1.3% of the strongest quasicrystal reflection, $02\bar{2}40\bar{4}$. The intensities of the other unidentified peaks are smaller than 0.3% of the $02\bar{2}40\bar{4}$ reflection. Referring to the quality of the quasicrystal specimen and the small concentration of an impurity phase, we regarded that the present Zn-Mg-Ho sample satisfied the prerequisite for the Rietveld analysis.

The structure model of the icosahedral quasicrystal was determined by the procedure described in the former section. Two models were obtained as final models, of which the parameters are summarized in Table 3. The two models resemble each other, and satisfy the conditions on the alloy composition and the mass density. The weighted reliability factor, $R_{wp}$, of Model B is slightly smaller than that of Model A, but the difference does not seem conclusive. In other words, it is difficult to select a unique structural model, based only on the results of the profile fitting. As an example, the calculated intensity distribution of Model B is presented in Fig. 1(b). The small difference between the measured and calculated distribution can be seen in Fig. 1(c). The difference is smaller in the high angle region rather than in the low angle region, as can be seen in Figs. 1(d) and (e). The measured and the calculated distributions are very similar with respect to the relatively strong reflections and also to the very weak reflections, for example $06\bar{4}606$ at $2\theta = 31.95°$. It is noted that more than 3000 reflections were treated in the calculation of the intensity distribution, but only a limited number of reflections, in the order of 400, contribute to the total intensity. This fact may indicate the special feature of the structure analysis of a quasicrystal, which is different from that of ordinary crystals.

To examine which model is appropriate, the occurrence of short interatomic distances was checked in the 3-dimensional atomic models generated in spherical regions with a radius of 100 Å. The number of atom pairs is presented in Fig. 2(a) and (b) in the cases of Models A and B, respectively. Recalling the presence of 2.54 Å between Zn and Zn in the $MgZn_2$ Laves phase, and 2.60 Å between Mg and Zn in the $Mg_{51}Zn_{20}$ phase, the interatomic distance, 2.60 Å, is considered to be the limit. Then, the interatomic distances, 0.61, 1.04 , 1.80 and 1.98Å appearing in both models are too short. In both models, the number of such unphysical short distance is very few as is seen in Fig. 2. The main difference of these two models can be noticed at the interatomic distance 3.10 Å, which is the distance between BC1 and EC, and between BC2 and EC. In Model A the number of this pair is approximately 0.7 times smaller than that in Model B. This is mainly due to the small radius of BC1 in Model A. Considering that there is almost no difference in the appearance of the unphysical short distances in these two models, it is concluded that Model A includes many holes or vacancies in its structure, and that Model B is appropriate as the starting model for future refinement. The choice of Model B seems not inconsistent with the result derived by the low density elimination method by Takakura et al. [6], although our former models [4,5] are quite similar to Model A.

The most remarkable feature of Model B is the presence of an almost complete 3-dimensional Penrose tiling formed by Node1 and Node2. The edge length of the Penrose tiling is estimated to be 5.20 Å from the 6-dimensional lattice parameter, $a_{6D}$=14.693 Å. The triacontahedron window [10,11] for the ideal Penrose tiling can be approximated by the sphere of radius 7.44 Å, which is approximately equal to the estimated radii of Node1 and Node2 in Model B.   Two kinds of Penrose rhombohedral tiles are presented in Fig. 3, which include the typical decoration by atoms.   The vertices of these rhombohedra are occupied by Zn (Node1) and Mg (Node2), alternatively, in accordance with the F-type ordering.   The other interesting feature is the presence of EC atoms at the edge-center positions of the Penrose tiles.   The window necessary for the ideal edge-center decoration is the rhombic icosahedron [12] with edge length 5.20 Å, which has a height of 11.62 Å along the 5-fold axis and a volume of 863 Å$^3$.   The ellipsoid window of Model B has a height of 9.18 Å and a volume of 868 Å$^3$. Then, this window approximates the ideal one.   The Ho and Mg atoms belonging to BC2 and BC1 respectively, seem to form another 3-dimensional Penrose tiling.   They tend to appear at two positions on the long body-diagonal of the prolate rhombohedron in Fig. 3(a), which divides the body-diagonal in the ratio of 1 to $\tau$ ($=(1+\sqrt{5})/2$).   This structure model of Zn-Mg-Ho quasicrystal is quite similar to that of Al-Li-Cu proposed by Elswijk et al. [15], and indicates that the Zn-Mg-Ho quasicrystal belongs to the Bergman type [4].   It should be noted that the properties of the 3-dimensional structure described here are somewhat idealized and are not strictly realized due to the use of spherical and ellipsoidal windows.   Although the present model includes few cases of too short distances, we believe that this model, Model B, contains the essential structural properties of the Zn-Mg-Ho icosahedral quasicrystal, and will be useful as a starting model to consider polyhedral windows in the future analysis.

   The large magnitudes of $B_\perp$, 54.0 Å$^2$, were estimated for the present structure model.   This value indicates that the mean square value of fluctuation $u_\perp$ in the perpendicular space is approximately 0.68 Å$^2$ .   This is not so small comparing with the dimension of the windows, and indicates the limitation of the present treatment.   In other words, we should argue the present result in this limit.

5. Conclusion

   Powder X-ray diffraction pattern of the Zn-Mg-Ho icosahedral quasicrystal was analyzed by the modified Rietveld method.   The result indicates the presence of the 3-dimensional Penrose tiling with edge-center decoration as a basic structure.   Atomic ordering related to F-type translational order is confirmed to be present in this quasicrystal, with Zn at Node1, Mg at Node2, Mg at BC1 and Ho at BC2.  However, it was difficult to obtain a unique structural model only by the analysis of X-ray diffraction pattern.  Furthermore, a large parameter of $B_\perp$ was estimated.  These facts may indicate the limitation of the present analysis using the simplified window model.

Acknowledgment

This work was supported by a Grant-in-Aid for Scientific Research B from Japan Society for the Promotion of Science.  The synchrotron radiation experiments were performed at SPring-8 BL02B2 with the approval of the Japan Synchrotron Radiation Research Institute.

Figure captions

Fig. 1    Powder X-ray diffraction patterns of Zn-Mg-Ho icosahedral quasicrystal: (a) measured pattern;    (b) calculated pattern of Model B;    (c) difference between the measured and calculated patterns.    (d) and (e) are magnified parts of (a) and (b), respectively.

Fig. 2    Histograms of interatomic distances appearing in (a) Model A and (b) Mode B.    The origin of some short distances is summarized in inset.

Fig. 3    Local atomic arrangements in Model B: (a) prolate rhombohedron with Mg and Ho atoms on its body diagonal with the edge length of 5.20 Å; (b) oblate rhombohedron.    The color denotes the atom type.

Table 1

6-dimensional structure model of the Zn-Mg-Ho icosahedral quasicrystal. The ellipsoidal window at EC has a short axis along the 5-fold direction.

| Site | Window | Representative | Order | Symmetry |
|------|--------|----------------|-------|----------|
| Node1 | sphere | [0 0 0 0 0 0] | 1 | Yh |
| Node2 | sphere | 1/2[1 0 0 0 0 0] | 1 | Yh |
| BC1 | sphere | 1/4[1 1 1 1 1 1] | 1 | Yh |
| BC2 | sphere | 1/4[3 1 1 1 1 1] | 1 | Yh |
| EC | ellipsoid | 1/4[1 0 0 0 0 0] | 12 | C5v |

Table 2

Fitting parameters used for the modified Rietveld analysis.   Concentric spherical shells are available for Node1 and Node2 as well as BC1 and BC2.

---

| Parameters | Meaning of parameters |
|---|---|
| $RN11, RN12$ [Å] | Radii of concentric spheres at Node1 |
| $RN21, RN22$ [Å] | Radii of concentric spheres at Node2 |
| $RB11 \sim RB13$ [Å] | Radii of concentric spheres at BC1 |
| $RB21 \sim RB23$ [Å] | Radii of concentric spheres at BC2 |
| $aEC, bEC$ [Å] | Radii of circle of ellipsoid located at EC.   $bEC$: along center axis. |
| $Shift$ [Å] | Shift of EC in the perpendicular space in the direction of the center axis.   Positive value means shift in the side of BC1. |
| $B_{//}$ [Å$^2$] | $B$ parameter in Debye-Waller factor in the physical space |
| $B_{\perp}$ [Å$^2$] | $B$ parameter in Debye-Waller factor in the perpendicular space |
| $a_{6D}$ [Å] | 6-dimensional lattice parameter |
| $S$ | Scale factor |
| $Z, D, T$ | Coefficients for correction of peak position $2\theta_k$ <br> $\Delta 2\theta_k = Z + D \cos\theta_k + T \sin^2\theta_k$ |
| $U, V, W$ | Coefficients for peak width Hk, full width at half maximum <br> $H_k = (U \tan^2\theta_k + V \tan\theta_k + W)^{1/2}$ |
| $A_0, A_1, A_2$ | Coefficients for asymmetry A of peak shape <br> $A = A_0 + A_1 (2^{1/2} - 1/\sin\theta_k) + A_2 (2 - 1/\sin^2\theta_k)$ |
| $\eta_{H0}, \eta_{H1}$ | $\eta$ parameters for pseudo Voigt function for high angle side <br> $\eta_H = \eta_{H0} + 2\theta_k \eta_{H1}$ |
| $\eta_{L0}, \eta_{L1}$ | $\eta$ parameters for pseudo Voigt function for low angle side <br> $\eta_L = \eta_{L0} + 2\theta_k \eta_{L1}$ |
| $BG_0 \sim BG_5$ | Coefficients of background function expanded by Legendre's orthogonal polynomials up to 5th order. |

---

Table 3

Two structure models for Zn-Mg-Ho icosahedral quasicrystals. The numbers in the parentheses are standard deviations in the Rietveld fitting.

| | Model A | Model B |
|---|---|---|
| $RN11$ [Å] | Zn: 7.389(4) | Zn: 7.525(4) |
| $RN21$ [Å] | Mg: 7.383(9) | Mg: 7.14(1) |
| $RB11$ [Å] | Zn: 5.047(6) | Mg: 7.50(2) |
| $RB21$ [Å] | Ho: 7.278(3) | Ho: 7.419(3) |
| EC | 0.68Zn+0.32Mg | 0.80Zn+0.20Mg |
| $aEC$ [Å] | 6.987(3) | 6.722(4) |
| $bEC$ [Å] | 4.764(6) | 4.588(7) |
| $Shift$ [Å] | 0.542(3) | 0.619(4) |
| $B_{//}$ [Å$^2$] | 1.614(5) | 1.519(5) |
| $B_{\perp}$ [Å$^2$] | 53.2(4) | 54.0(4) |
| $a_{6D}$ [Å] | 14.6935(1) | 14.6933(1) |
| $R_{wp}$ [%] | 6.1 | 5.8 |
| Mg [at.%] | 31.5 | 31.3 |
| Zn [at.%] | 59.1 | 58.8 |
| Ho [at.%] | 9.4 | 9.9 |
| $\rho$ [g/cm$^3$] | 5.62 | 5.68 |

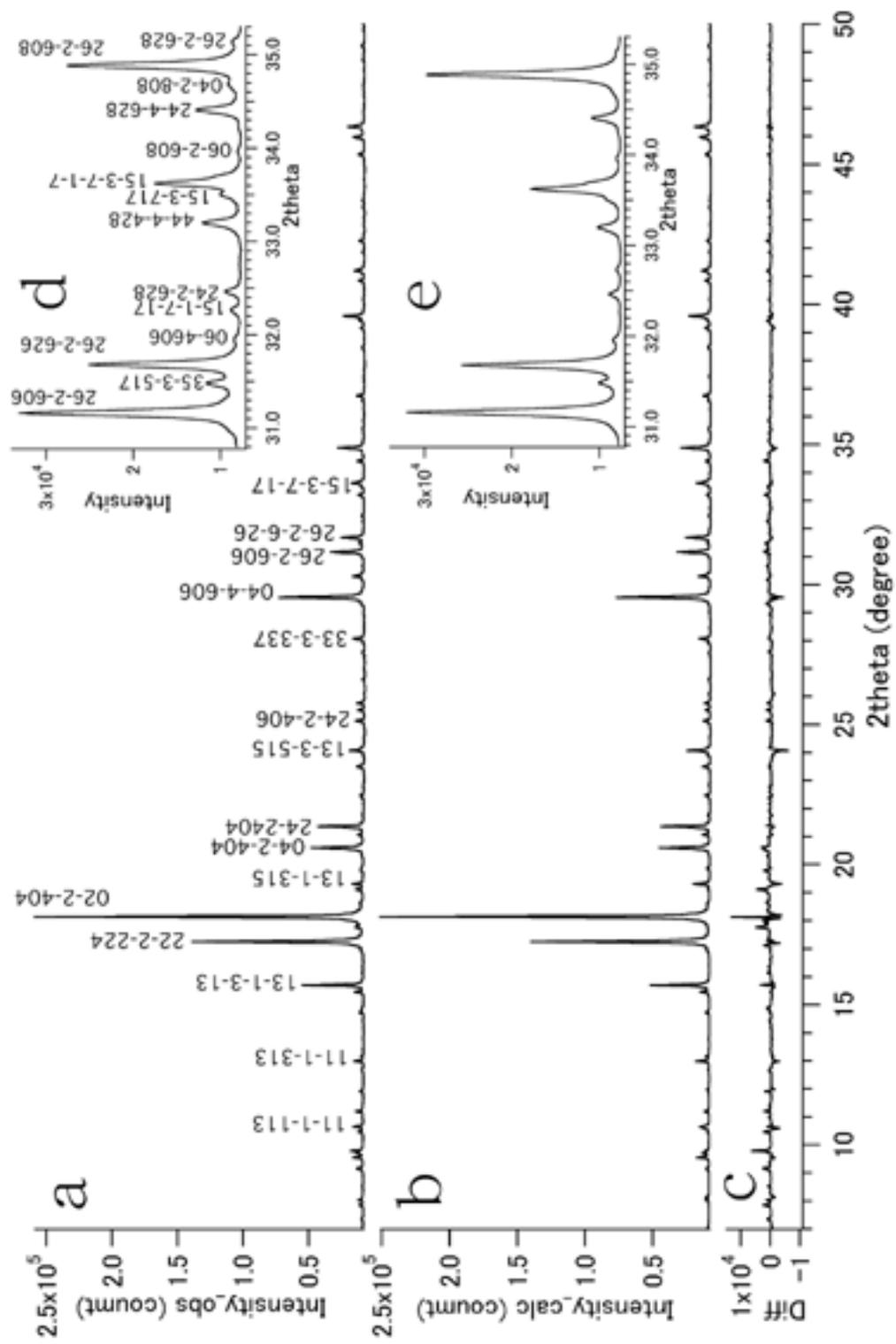

Fig. 1

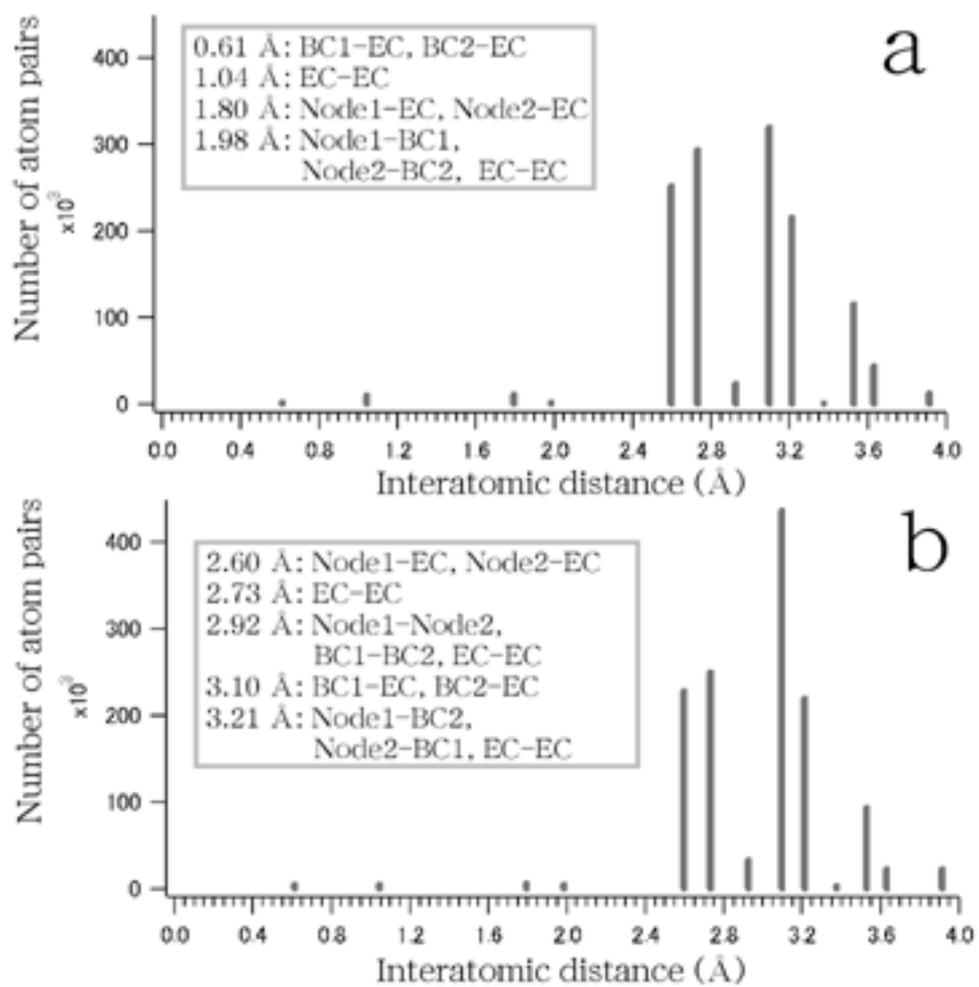

Fig. 2

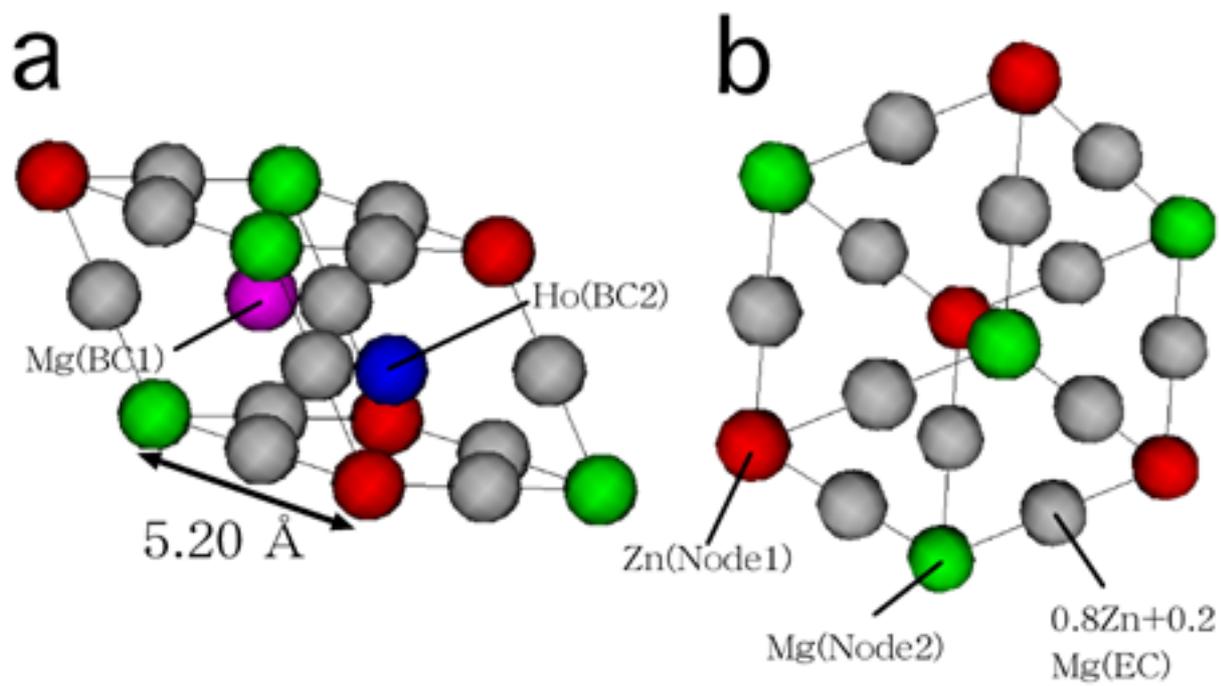

Fig. 3